\documentclass[12pt,a4paper]{article}

\usepackage{cite}

\usepackage{amssymb,amsmath}

\newcommand{\al}{\alpha}
\newcommand{\be}{\beta}

\newcommand{\st}{\ast\!}

\newcommand{\ve}{\varepsilon}

\newcommand{\ba}{\begin{eqnarray}}
\newcommand{\ea}{\end{eqnarray}}
\newcommand{\bpm}{\begin{array}}
\newcommand{\epm}{\end{array}}

\begin{document}


{\bf \centerline{\Large D-dimensional metrics with D-3 symmetries}}

\bigskip

\centerline{ A Szereszewski, J Tafel and M Jakimowicz}

\bigskip
\noindent
Institute of Theoretical Physics, Faculty of Physics, University of Warsaw,  Poland

\bigskip
\noindent
E-mail: aszer@fuw.edu.pl

\begin{abstract}
Hidden symmetry transformations of $D$-dimensional vacuum metrics with $D-3$ commuting Killing vectors are studied. We solve directly the Einstein equations in the Maison formulation under additional assumptions. We relate the 4-dimensional Reissner-Nordstr\"om solution to  a particular  case of the 5-dimensional Gross-Perry metric.

\end{abstract}

\section{Introduction}

Stationary vacuum Einstein equations admit the symmetry group $SL(2,\mathbb{R})$ \cite{E,NK}. This symmetry was generalized to an action of the group $SL(D-2,\mathbb{R})$ in a class of $D$-dimensional vacuum metrics with $D-3$ commuting Killing vectors \cite{M}. In the case $D=5$ this group contains the group $SO(1,2)$ which preserves asymptotical flatness of a metric \cite{GS}. For instance, using this action one can reproduce the Myers-Perry solution from the Schwarzschild-Tangherlini metric \cite{GS}.

In this paper we investigate in detail the action of the $SL(D-2,\mathbb{R})$ symmetry group when integral submanifolds of the Killing vectors are  spacelike or timelike. We identify relevant parameters of this action and we disscuss corresponding changes of the metric signature. We solve directly the Einstein equations in $D=5$ assuming two commuting Killing vectors and additional  symmetries which are not isometries. 
We give an example of the $SL(3,\mathbb{R})$ symmetry transformation   which generates the Reissner-Nordstr\"om 4-dimensional solution, with a dyonic electromagnetic field, from the 5-dimensional Gross-Perry metric \cite{GP}  of the Euclidean signature.

In the considered class of solutions there are near horizon metrics of extremal black holes \cite{HI} and
metrics obtained by Cl\'ement (see \cite{C3} and references therein).

\section{Generation method for reduced vacuum Einstein equations}
Let $M$ be  $(n+2)$-dimensional manifold with  metric $g$ admitting $n-1$ commuting Killing vectors which define a non-null integrable distribution. In special coordinates
$x^i$, $i=1,\dots,n-1$, and $x^a$,  $a=n,n+1,n+2$, the Killing vectors are $\partial_i$ and the metric takes the form
				 \ba
g=g_{ij}(d x^i+A^i)(d x^j+A^j)+\tau^{-1}\tilde g_{ab}d x^a d x^b , \label{met}
				 \ea
where $\tau=|\det g_{ij}|$, $A^i=A^i_{\ a} dx^a$ and functions $g_{ij}$, $A^i_{\ a}$, $\tilde g_{ab}$ do not depend on coordinates $x^i$.
	The vacuum Einstein equations for metrics (\ref{met}) are equivalent to the following equations
				 \ba
	      &d (\chi^{-1}\st d\chi)=0, \label{eqn1}\\
&\tilde R_{ab}=\frac{1}{4}\textrm{Tr} (\chi^{-1}\chi_{,a}\chi^{-1}\chi_{,b}) \label{eqn2}
				 \ea
				 for $\tilde g_{ab}$ and $n\times n$ symmetric matrix $\chi$ constrained by the conditions
				 \ba\label{2a}
				 \det{\chi}=\pm 1\ ,\ \ \ \ \ \ve\chi_{nn}<0\ ,
				 \ea
where	$\ve=\textrm{sgn}(\det \tilde g_{ab})$. 	Here  $\tilde R_{ab}$ is the Ricci tensor of the metric $\tilde g=\tilde g_{ab} d x^a d x^b$ and 
$*$ denotes the Hodge dualization with respect to this metric.		 
				 The matrix $\chi$ is related to components  of  (\ref{met}) via the equations
				 \ba
		     \chi=\left(\begin{array}{cc}
  g_{ij}-\frac{\ve}{\tau}V_iV_j  &  \frac{1}{\tau}V_i \\
      \frac{1}{\tau}V_j     &  -\frac{\ve}{\tau} \label{chi}
        \end{array}\right) ,
 \ea
 \ba
 dV_i=\tau g_{ij}\st d A^j\ .\label{3}
 \ea
 Note that equation (\ref{3}) is integrable by virtue of (\ref{eqn1}) and that 
 \ba
 \textrm{sgn}(\det g_{ij})=-\ve \det{\chi}\ .
 \ea

Equations (\ref{eqn1})-(\ref{2a}) are preserved by the transformation
 \ba\label{sym}
    \chi \mapsto \chi'=\ve' S^T\chi S
    \ea
    where $S\in SL(n,\mathbb{R})$ is a constant matrix and value of $\ve'=\pm 1$ is fixed by the condition
 \ba
 \ve'\ve(S^T\chi S)_{nn}<0\ .\label{5}
 \ea
Transformations given by (\ref{sym})-(\ref{5})  can be used to obtain new vacuum metrics from known ones.
 They  generalize the Ehlers transformation \cite{E} for stationary 4-dimensional metrics. In dimension 5 they contain the group $SO(1,2)$ preserving asymptotical flatness of metrics \cite{GS}.
  
\section{Relevant parameters  and change of signature}
 
 Most of  parameters in $S$ (symmetry) do not change the seed metric in a nontrivial way. 
Any  matrix $S$ with $S^n_{\ n}\neq 0$ can be uniquely decomposed into a product of three matrices
$S=S_0HT$, where
 \ba 
  S_0&=&\left(\begin{array}{cc}
     \delta^i\,_j & \alpha^i \\ 
             0      &   1  
   \end{array}\right)\ ,\\
  H&=&\left(\begin{array}{cc} 
      \beta^i\,_{j}   &    0 \\
         0            \ \ & \left(\det \beta^i\,_{j}\right)^{-1}  \end{array}\right) , \quad
  T=\left(\begin{array}{cc}
      \delta^i\,_{j}   &    0 \\
         \gamma_j        &   1  \end{array}\right) .    \label{NRL}
 \ea 
The matrix $T$ yields translations of $V_i $ by  constants  $-\ve\gamma_i$. Its action on a seed metric is trivial. The matrix $H$ corresponds
to a linear transformation of $x^i$ and $A^i$ combined with a multiplication of the full metric $g$ by $(\det \beta^i\,_{j})^{-2}$. Thus, modulo  coordinate transformations, $H$ is a homothety and can be replaced by 
\ba
H_0=\left(\begin{array}{cc} 
     \beta \delta^i\,_{j}   &    0 \\
         0        &   \beta^{1-n} \end{array}\right)
         \ea
         with an appropriate constant $\beta$. 
The only nontrivial action is that of the matrix $S_0$ (true symmetry), 
 \ba\label{N}
    S_0^T\chi S_0=\left(\begin{array}{cc}
   g_{ij}-\frac{\ve}{\tau}V_i V_j  &  \alpha_i-\frac{\ve}{\tau}V_i\left(\alpha^kV_k-\ve\right)\\
  \alpha_j-\frac{\ve}{\tau}V_j\left(\alpha^kV_k-\ve\right) & V^k V_k-
                      \frac{\ve}{\tau}\left(\alpha^kV_k-\ve\right)^2  \end{array}\right),
\ea
where $\alpha_i=g_{ij}\alpha^j$ and $V^i=g^{ij}V_j$.

If $S^n_{\ n}=0$ then  $S$ is equivalent, modulo $H$ and $T$, to one of the matrices $S_l$, $l=1\dots n-1$, given by
\begin{equation}
 \begin{split}\label{mk}
  S^i_{ln}=&\left\{\begin{array}{lcl}
           1 &\textrm{if}& i= n-l\\
           0 &\textrm{if}& i> n-l
                 \end{array}\right. \ ,\ 
               S^n_{lj}=\left\{\begin{array}{lcl}
           -1 &\textrm{if}& j= n-l\\
           0 & \textrm{if}& j\neq n-l   
                        \end{array}\right. \ ,\\
  S^i_{lj}=&\left\{\begin{array}{lcl}
           1 &\textrm{if}& i=j\neq n-l\\
           0 & \textrm{if}& i\neq j\ \textrm{or}\ i=j= n-l             
                             \end{array}\right. \ .
 \end{split}
\end{equation}
Note that $S_l$ contains $n-l-1$ free parameters (components $S^i_{ln}$ for $i< n-l$). 
Summarizing, without loss of generality, symmetry transformations (\ref{sym}) can be reduced to the action of one of matrices $S_0,S_l$ composed with $H_0$, the latter equivalent to a homothety of $g$.

Transformations (\ref{sym}) can change signature of  $g_{ij}$, and hence the signature of (\ref{met}). Let the initial  signature of $g_{ij}$   be $(p,q)$, where $p$ denotes the multiplicity of the value +1. If $\ve'>0$ then  transformation (\ref{sym}) preserves this signature.
If $\ve'<0$ then the signature of the final metric $g_{ij}$  is $(q+\ve,p-\ve)$. Note that if $(p,\ve)=(0,1)$ or $(q,\ve)=
(0,-1)$ then $\ve'>0$. For instance, if we start with a 5-dimensional metric of the Lorentz signature $(-++++)$ then the transformed metric has the same signature or the Euclidean one.
Transformation (\ref{N}) adds two parameters to the seed metric.  (They become dependent, $\alpha^2=\frac{1}{2}(\alpha^1)^2$, if the asymptotical flatness is to be preserved \cite{GS}.) Transformations $S_l$ take the form
\ba
   S_1=\left(\begin{array}{ccc}
        1 & 0 & \alpha \\
        0 & 0 & 1\\
        0 & -1 & 0
      \end{array}\right)\ ,
  \ea
\ba
     S_2=\left(\begin{array}{ccc}
        0 & 0 & 1 \\
        0 & 1 & 0 \\
        -1 & 0 & 0
      \end{array}\right)\ .
  \ea

\section{Solutions with 2-dimensional space of constant curvature}

Let us  assume that $\chi=\chi(z)$ depends only on one coordinate $z$ and metric  $\tilde{g}$ has the following form
 \ba
   \tilde{g}=dz^2+f(z){g}^{(k)}, \label{hmet}
 \ea
where ${g}^{(k)}$ is a 2-dimensional metric of constant curvature $k=0,\pm 1$ and signature $++$ or $+-$,
 \ba
   {g}^{(k)}=\frac{4\left(d\tilde x^2+\ve d\tilde y^2\right)}{\left(1+k( \tilde x^2+\ve \tilde y^2)\right)^2}\ .
 \ea 
 Note that $z$ can be shifted by a constant and metric (\ref{hmet}) can be multiplied by another constant since this transformation can be compensated by a change of coordinates $x^i$. Thus, $\tilde g$ can be simplified by means of transformations
 \ba\label{7}
 z\mapsto cz+c_0\ ,\quad f\mapsto c^{-2}f\ ,\quad c,c_0=\textrm{const}.
 \ea
 
 The Ricci tensor of $\tilde{g}$ reads
 \ba
\textrm{Ricci}(\tilde{g})=\frac{(f_{,z})^2-2ff_{,zz}}{2f^2}dz^2+\left(k-\frac{f_{,zz}}{2}\right){g}^{(k)}.
 \ea
It follows from (\ref{eqn2}) that 
\ba\label{6}
f=kz^2+az+b,\ \ a,b=\textrm{const}
\ea
and
\ba
 \textrm{Tr}\left(\chi^{-1}\chi_{,z}\right)^2=2a^2-8kb\ .\label{8}
   \ea
A double integration of (\ref{eqn1}) gives
 \ba
   \chi=\chi_0\exp\big((w+w_0)C\big),
 \ea
where  $\chi_0$, $C$ are constant matrices such that 
\ba
   \chi_0=\chi_0^T, \quad  \chi_0 C=C^T\chi_0, \quad \textrm{Tr}\,C=0, \quad  \textrm{Tr}\ C^2=2a^2-8kb
   \label{chi0C}
 \ea
 and $w(z)$ is a particular solution of
\ba\label{9}
w_{,z}=\pm f^{-1}\ .
 \ea
The constant $w_0$ and the sign in (\ref{9}) can be arbitrarily chosen.
 
One can  classify functions $f$ and $w$ by putting them into a canonical form. First, let us note that 
  by virtue of (\ref{7}) one can transform $f$ into one of the following  expressions labelled by  $k$ 
and a new index $k'=0,\pm 1$,
 \ba
 f^{(k,k')}=\left\{\begin{array}{lcl}
           z & \textrm{if} & k=k'=0\\
           kz^2+k' & \textrm{if} & k^2+k'^2\neq 0.   
               \end{array}\right. \label{f}
 \ea 
 Using (\ref{f}) we find particular solutions of (\ref{9}). They  are presented in the  table
\ba
\begin{tabular}{l|l|c|c}
     & $w(z)$                  &  conditions & Tr $C^2$  \\ \hline
   i)& $\log |z|$              &  $k=k'=0$  & 2  \\ \hline
  ii)& $\textrm{arctanh}\, z$  &  $k'=-k\neq 0,\ z^2<1$  & 8  \\ \hline
 iii)& $\textrm{arccoth}\, z$  &  $k'=-k\neq 0,\  z^2>1$  & 8  \\ \hline
  iv)& $z$                     &  $k=0,\ k'\neq 0$  & 0  \\ \hline
   v)& $1/z$                   &  $k\neq 0,\ k'=0$  & 0  \\ \hline  
  vi)& $\textrm{arccot}\, z$   &  $k'=k\neq 0$  & -8  \\ \hline
\end{tabular}\ . \label{w}
\ea

The symmetry (\ref{sym}) induces  transformations $C\mapsto S^{-1}CS$ and 
$\chi_0\mapsto S^{T}\chi_0 S$ which allow us to 
reduce matrices $C$ and $\chi_0$ to  simpler forms.  First, we put matrix
$C$ into a canonical Jordan form. Then, we find  $\chi_0$
satisfying (\ref{chi0C}) and we simplify it by means of  matrices which commute with $C$. Below we present results of this procedure in the case
of 5-dimensional metrics. Then $n=2$ and  there are four canonical forms of $C$ and $\chi$, in which 
$\alpha, \beta=\textrm{const}$ and $\ve$,$\ve_i=\pm 1$,
\begin{equation}
  \begin{split}  
   C&=\left(\begin{array}{ccc}
        \al & \beta & 0 \\
       -\beta & \al & 0\\
        0 & 0 & -2\al
      \end{array}\right),\  
   \chi=\left(\begin{array}{ccc}
       -e^{\al w}\sin(\beta w) &  e^{\al w}\cos(\beta w) & 0 \\
        e^{\al w}\cos(\beta w) &   e^{\al w}\sin(\beta w) & 0 \\
                       0                  &     0                 & -\ve e^{-2\al w} 
      \end{array}\right), \\
      &\textrm{Tr}\  C^2=6\al^2-2\beta^2,\ \beta\neq 0 .
 \end{split}\label{26}
\end{equation}
\begin{equation}
 \begin{split} 
     C&=\left(\begin{array}{ccc}
        \al &    0  & 0\\
         0  & \beta & 0\\
         0  &    0  & -\al-\beta 
           \end{array}\right), \ \    
   \chi=\left(\begin{array}{ccc}
        \ve_1e^{\al w}       &    0   & 0\\
         0         & \ve_2 e^{\beta w} & 0\\
         0         &    0   & -\ve e^{-(\al+\beta)w} 
           \end{array}\right),\\
           &\textrm{Tr}\ C^2=2(\al^2+\al\beta+\beta^2)\label{27}
           \end{split}
  \end{equation}
\begin{equation}
  \begin{split}     
     C&=\left(\begin{array}{ccc}
        \al & 1 & 0 \\
        0 & \al & 0\\
        0 & 0 & -2\al
      \end{array}\right),\ 
     \chi=\left(\begin{array}{ccc}
        0 & \ve_1 e^{\al w} & 0 \\
        \ve_1  e^{\al w} & \ve_1w e^{\al w} & 0\\
        0 & 0 & -\ve e^{-2\al w}
      \end{array}\right),\\ 
   &\textrm{Tr}\ C^2=6\al^2 
    \end{split}\label{28}
  \end{equation}
\begin{equation}
  \begin{split}
    C=\left(\begin{array}{ccc}
        0 & 1 & 0 \\
        0 & 0 & 1\\
        0 & 0 & 0
      \end{array}\right), \ 
     \chi&=-\ve\left(\begin{array}{ccc}
        0 &    0    & 1 \\
        0 &    1    & w\\
        1 &    w    & \frac{1}{2} w^2
      \end{array}\right),\
     \textrm{Tr}\ C^2=0 
 \end{split}\label{29}
\end{equation} 
Note that case (\ref{26}), with  parameters in an appropriate range, can be merged with any function from table  (\ref{w}), 
cases (\ref{27}) and (\ref{28}) admit  functions i)-v)  while (\ref{29}) is  only compatible with 
functions iv)-v).

\section{Example}
As an example we consider metrics related to  (\ref{27}) with $k=-k'\neq 0$. In this case we obtain
 \begin{multline}
 g=\ve_1 \left|\frac{z+1}{z-1}\right|^{\al/2}(dx^1)^2+
    \ve_2 \left|\frac{z+1}{z-1}\right|^{\be/2}(dx^2)^2\\
  +\left|\frac{z+1}{z-1}\right|^{-(\al+\be)/2}\left(dz^2+k(z^2-1)g^{(k)}\right), \label{i}
  \end{multline}
 where $\al^2+\al\be+\be^2=4$.  If  $\ve_1=\ve_2=\ve=k=1$, $\al=-\be=-2$ and we introduce a multiplicative constant in $g$ we obtain  the
Euclidean Gross-Perry solution
 \ba
   g=\left(\frac{\rho-q}{\rho+q}\right)^{2}(dx^1)^2+
     \left(\frac{\rho+q}{\rho-q}\right)^{2}(dx^2)^2
  +\frac{1}{4}\left(1-\frac{q^2}{\rho^2}\right)^2
    \left(d\rho^2+\rho^2 d\Omega^2\right),\   \label{i_rho}
 \ea
where $d\Omega^2=d\theta^2+\sin^2\theta d\phi^2$,  $q$=const and  
$\rho=q(z+\sqrt{z^2-1})$.
The metric (\ref{i_rho}) can be transformed by means of (\ref{sym}) with 
 \ba
    S=-\frac{1}{2q}\left(\begin{array}{ccc}
                  \sqrt{2(m^2-q^2)}   &   m+q                 &  m-q \\
                 -\sqrt{2(m^2-q^2)}   &  -m+q                 &  -m-q \\
                         2m           &  \sqrt{2(m^2-q^2)}    &  \sqrt{2(m^2-q^2)}
                \end{array}\right).
 \ea
For $\ve'=-1$ this transformation yields the following 5-dimensional vacuum solution
  \begin{multline} 
   g=\left(dx^1-\sqrt{2} Q \cos\theta d\phi + \frac{\sqrt{2}Q}{r} dt\right)^2-\left(1 -  \frac{2m}{r} + 
\frac{Q^2}{r^2}\right) dt^2\\
      + \left(1 -\frac{2m}{r} +\frac{Q^2}{r^2}\right)^{-1} dr^2 +   r^2 d\Omega^2,
 \end{multline}
where $t=x^2$, $\rho=r-m+\sqrt{r^2-2mr+Q^2}$ and  $q=\sqrt{m^2-Q^2}$. Within the Kaluza-Klein approach the latter metric decomposes into
the 4-dimensional Reissner-Nordstr\"om solution
 \ba
   g'=-\left(1 -  \frac{2m}{r} + 
\frac{Q^2}{r^2}\right) dt^2
      + \left(1 -\frac{2m}{r} +\frac{Q^2}{r^2}\right)^{-1} dr^2 +   r^2 d\Omega^2
 \ea
and  electromagnetic field given by 
\ba
A=-\sqrt{2} Q \cos\theta d\phi + \frac{\sqrt{2}Q}{r} dt\ .
\ea
This field represents a magnetic and electric monopoles of the same strength and placed at the same point.

\section*{Acknowledgments}
We are grateful to  G. Cl\'ement for drawing our attention to \cite{C3} and 
references therein.

This work is partially supported by the grant N N202 104838 of Polish Ministerstwo Nauki i Szkolnictwa Wy\.zszego.

\end{document}